\documentclass[aps,prl,twocolumn,showpacs]{revtex4}
\usepackage{graphicx,amsmath,amssymb,color}
\usepackage[normalem]{ulem}
\usepackage{amsfonts}
\usepackage[toc,page]{appendix}
\usepackage[ocgcolorlinks,colorlinks=true,linkcolor=blue,citecolor=red]{hyperref}
\usepackage{latexsym}
\usepackage{amsfonts}
\usepackage{algpseudocode}
\usepackage{amsthm}
\usepackage{mathrsfs}
\usepackage{natbib}
\usepackage{color,verbatim}

\def\beq{\begin{eqnarray}}
\def\eeq{\end{eqnarray}}

\begin{document}
\date{\today}
\addtolength{\topmargin}{0.75in}

\bibliographystyle{unsrt}
\title{Circuit design for multi-body interactions in superconducting quantum annealing systems with applications to a scalable architecture}

\author{N.\ Chancellor$^{\dagger,1, \star}$, S.\ Zohren$^{\dagger,2}$ and P.\  A.\ Warburton$^{1,3}$ }

\affiliation{$^1$London Centre for Nanotechnology 19 Gordon St, London, UK  \\
$^2$ Department of Materials and Department of Engineering Science, University of Oxford, Parks Road, Oxford, UK \\
$^3$Department of Electronic and Electrical Engineering, UCL, Torrington Place, London, UK \\
$^\star$ Current Affiliation: Department of Physics, Durham University, South Road, Durham, UK  \\
$^\dagger$Authors contributed equally.
}


\date{\today}

\maketitle

{\bf Quantum annealing provides a way of solving optimization problems by encoding them as Ising spin models which are implemented using physical qubits. The solution of the optimization problem then corresponds to the ground state of the system. Quantum tunnelling is harnessed to enable the system to move to the ground state in a potentially highly non-convex energy landscape. A major difficulty in encoding optimization problems in physical quantum annealing devices is the fact that many real world optimization problems require interactions of higher connectivity as well as multi-body terms beyond the limitations of the physical hardware. In this work we address the question of how to implement multi-body interactions using hardware which natively only provides two-body interactions. The main result is an efficient circuit design of such multi-body terms using superconducting flux qubits in which effective N-body interactions are implemented using N ancilla qubits and only two inductive couplers. It is then shown how this circuit can be used as the unit cell of a scalable architecture by applying it to a recently proposed embedding technique for constructing an architecture of logical qubits with arbitrary connectivity using physical qubits which have nearest-neighbor four-body interactions.  It is further shown that this design is robust to non-linear effects in the coupling loops as well as mismatches in some of the circuit parameters.}

\section{Introduction}

Solving machine learning and optimization problems by casting them as an Ising spin glass and then using a physical device to take advantage of quantum fluctuations has been a subject of much recent interest \cite{OGorman2015, Neven2008, Santra2014, Hen2012, Perdomo-Ortiz2012, Boixo2014, Venturelli2014, Reiffel2015, Chancellor2015,Benedetti2016,Benedetti2016a}. This interest is due in a large part to demonstration of the underlying principles of quantum annealing in condensed matter systems \cite{Brooke1999} and the more recent development of a programmable annealing device by D-Wave Systems Inc. \cite{Johnson2011, Harris2010}. 

Although the niobium SQUIDs which are the basic building blocks of the D-Wave annealer display limited coherence, it has been used to demonstrate that quantum tunnelling is an exploitable resource in a computational setting \cite{Denchev2016}. Furthermore the development of annealers using aluminium SQUIDs with orders-of-magnitude longer coherence lifetimes \cite{Fan2016} may enable further improvements in the computational performance of future quantum annealers.

Mapping real world problems, or indeed problems with a similar difficulty to interesting real world problems, to a programmable annealer is a major practical challenge \cite{Hen2015, Katzgraber2015}. It is known for example that even though finding the ground state of the native so-called Chimera graph of the D-Wave device is NP-hard, typical randomly generated instances on this graph are actually easy to solve by simulated annealing type algorithms  \cite{Katzgraber2014}. Because it is non-planar, minor embedding can be used to map a fully connected graph to the Chimera, although at a significant overhead \cite{Choi2011, Venturelli2014}. 

The NP-completeness of finding the ground state of an arbitrary (2-body) Ising spin glass and therefore a Chimera graph guarantees that any NP-complete problem can be mapped to finding the ground state of a Hamiltonian which is a subgraph of a Chimera with polynomial overhead. For examples of how this can be done in practice, see \cite{Choi2004, Zhengbing2014,Biamonte2008,Whitfield2012}. However, there is no indication that this approach is optimal. For problems which require higher order interactions a mapping must be found from a Higher Order Binary Optimization (HOBO) to a Quadratic Unconstrained Binary Optimization problem (QUBO) \cite{Zhengbing2014}. Typically this is done iteratively, with an N-body interaction being reduced to a 2-body interaction using a complete graph on $N$ logical bits and $(N-2)$ ancilla bits \cite{Perdomo-Ortiz2008}. This necessitates $(N-1)(2\,N-3)$ two-body couplers.

In this letter, and a related work \cite{Chancellor/Zohren2015} we examine an alternative architecture, in which higher order problems can be expressed natively by coupling logical qubits to a group of ancillae. This letter focuses on native circuit implementations of this architecture, while \cite{Chancellor/Zohren2015} examines how the same principles can be applied within the Chimera architecture. 

Implementing a single logical clause using the methods of \cite{Choi2004, Zhengbing2014}, would require the construction of a penalty function (up to an unimportant energy offset) on a set of logical qubits, where the penalty is 0 if the clause is satisfied and greater than or equal to a penalty weight $g$ if the clause is violated. In the ideal case $g$ should be infinite, but in practice its maximal value is limited by the hardware.


This function gives a generally different penalty to all ``wrong'' answers (i.e. the ones which violate the clause). Hence the ground state of sums of more than one of these penalty functions (i.e. a sum of multiple penalty functions on overlapping subsets of bits) is meaningful if there exists a solution which violates zero or only one of the clauses. In the 3-SAT example in \cite{Choi2004}, this kind of superposition is acceptable because the problem is cast as a decision problem of whether a state exists which satisfies all of the penalties. However under a simple generalization of the problem to max-3-SAT, where we ask what is the choice which violates the least number of the constraints, this kind of superposition no longer yields a valid expression of the problem. 

Our proposal on the other hand is to construct a function which reproduces the spectrum of a high order penalty term, which is equal to zero if the clause is satisfied and exactly equal to the penalty weight $g$ if the clause is violated. Because all states which violate the clause are penalized equally, the ground state of a sum of an arbitrary number of such terms will in fact be the state which violates the smallest number of clauses, regardless of how many can be simultaneously satisfied. In general it may also be interesting to penalize different violations differently, (i.e. weighted max-k-SAT) but in a controlled way; our method also supports this.

As an example of how these techniques can be used, we explicitly show how to construct superconducting circuits which realize the fully connected architecture recently proposed in \cite{Lechner2015} as an alternative to the Chimera graph with minor embedding. We show that this architecture allows us an additional freedom in choosing the annealing path which is not a feature in the current D-Wave device architecture. While a recent numerical study \cite{Albash2016} has cast doubt on whether this method of embedding problems will perform better than the method proposed in \cite{Choi2011}, it does still have some novel features such as a greater richness of potential decoding methods.

For simplicity, and because it is what is required for the architecture in \cite{Lechner2015}, we will restrict ourselves to discussing how to reproduce the classical spectrum of multi-body operators of the form $\mathcal{H}_N= J_N \sigma^{z}_1 \ldots \sigma^{z}_N$. Such multi-body terms are important for many applications. For example, it is known that spin glasses undergo a transition in complexity when moving from 2-body couplings to multi-body terms of order 3 and higher \cite{Sherrington,Auffinger,Thomas2000}. In this transition the number of extrema of the energy landscape growths from a polynomial to an exponential function in the number of spins, where at the same time a banded structure in energy appears for saddle points of various orders \cite{Auffinger}. The latter has various implications for example for energy landscapes of deep neural networks which are related to spin glasses and where the order of the multi-body term of the spin glass is given by the depth of the network \cite{LeCunn1,LeCunn2}. This transition is in terms of the typical energy landscape structure. The fact that the 2-local Ising model is universal in the sense of being able to simulate classical Hamiltonians \cite{De-las-Cuevas2016} means that it could mimic the landscape of any Hamiltonian.

\section{results}

As discussed above, in many foreseeable applications for adiabatic quantum computation, one needs implementations of interactions with multi-body terms. These interactions are of the type
\beq \label{HN}
\mathcal{H}_N= J_N \sigma^{z}_1 \ldots \sigma^{z}_N.
\eeq 
However, the architecture used in quantum annealing devices generally only allows for two-body interactions, leading to Hamiltonians of the form
\beq \label{H2local}
\mathcal{H}= \sum_{i\sim j}   J_{ij } \sigma^{z}_i \sigma^{z}_j + \sum_i  h_i \sigma^{z}_i
\eeq 
where the first sum is taken over all adjacent qubits in the connectivity graph of the architecture.

We now describe a construction which reproduces the low-energy spectrum of \eqref{HN} using a Hamiltonian with two-body interactions of the form of \eqref{H2local} including additional ancilla qubits. This is first done on a theoretical level. In the next section we present an efficient circuit design of this construction. 

\begin{figure}
\begin{centering}
\includegraphics[width=6cm]{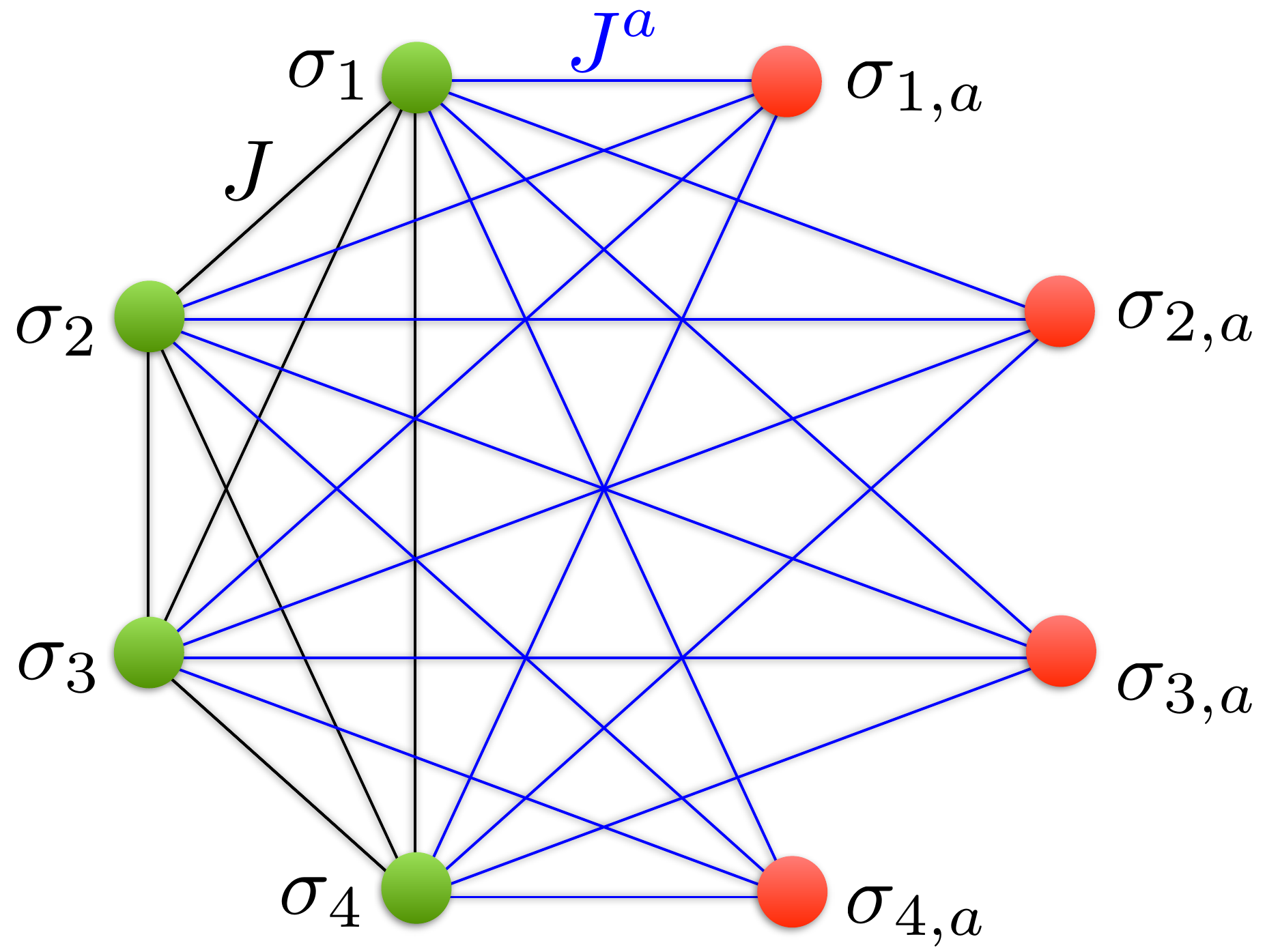}
\par\end{centering}
\caption{\label{fig1} Graph showing the connectivity of the Hamiltonian \eqref{H2localemb} for $N=4$. The green vertices represent the logical qubits while the red vertices represent the ancillas. The logical qubits are fully connected amongst themselves with two-body couplings, represented as edges, of strength $J$ (black). Each ancilla is connected to every logical qubit with two-body couplings of strength $J^a$ (blue).}
\end{figure}

\begin{table}[t]
\begin{center}
\begin{tabular}{| l | c |  c |}
\hline
Logical qubit states & Ancilla state  & E  \\
\hline 
$\uparrow \uparrow \uparrow \uparrow $ & $\downarrow \downarrow \downarrow \downarrow$  & $+J_4$  \\
$\downarrow \uparrow \uparrow \uparrow$, $ \uparrow \downarrow \uparrow \uparrow$, $\uparrow \uparrow \downarrow\uparrow$, $\uparrow \uparrow \uparrow \downarrow$
  & $\uparrow\downarrow \downarrow \downarrow$ & $-J_4$  \\
$\downarrow\downarrow\uparrow\uparrow$, $\downarrow \uparrow \downarrow \uparrow$, $\downarrow \uparrow  \uparrow \downarrow$, $\uparrow \downarrow\downarrow \uparrow$,
$\uparrow\downarrow \uparrow \downarrow\uparrow$, $\uparrow\uparrow\downarrow\downarrow$
  &  $\uparrow \uparrow\downarrow \downarrow$  & $+J_4$ \\ 
$\uparrow \downarrow \downarrow \downarrow$, $ \downarrow \uparrow \downarrow \downarrow$, $\downarrow \downarrow \uparrow \downarrow$, $\downarrow \downarrow \downarrow \uparrow$ 
   & $\uparrow \uparrow \uparrow\downarrow$ & $-J_4$  \\
$\downarrow \downarrow \downarrow \downarrow$  & $\uparrow \uparrow \uparrow \uparrow $ & $+J_4$ \\
\hline
\end{tabular}
\end{center}
\label{table}
\caption{Illustration of the various states of the 4-body term $\mathcal{H}_4= J_4 \sigma^{z}_1 \sigma^{z}_2\sigma^{z}_3 \sigma^{z}_4$. Shown are the states of the logical qubits, the minimum energy states of the ancillae and the total energy.}
\end{table}%

The Hamiltonian which reproduces the spectrum of the $N$-body term in \eqref{HN} constitutes $N$ logical qubits which are fully connected and an additional $N$ ancilla qubits which are connected to all logical qubits but not amongst each other. This is illustrated in Figure \ref{fig1} for the case of $N=4$. The reason behind this construction will become apparent below. However, we can already deduce that if this construction is to reproduce the low energy spectrum of \eqref{HN}, then by symmetry the logical qubits must all have equal magnetic fields, $h$, as well as equal two-body couplings, $J$, amongst each other. The same is true for the two-body couplings between the logical qubits and the ancillas, here denoted by $J^a$. This leads to the following Hamiltonian
\begin{eqnarray}
\mathcal{H}_N^{(2)}&=& J \sum_{i=2}^N \sum_{j =1}^{i-1} \sigma^{z}_i \sigma^{z}_j + h \sum_{i=1}^N  \sigma^{z}_i + \nonumber \\ 
&& \quad + J^a \sum_{i=1}^N \sum_{j=1}^N \sigma^{z}_i \sigma^{z}_{j,a} +   \sum_{i=1}^N  h_i^a \sigma^{z}_{i,a}. 
\label{H2localemb}
\end{eqnarray}

This construction relies on symmetry to effectively count the number of logical bits in the up orientation. By symmetry, the effect of the logical bits on the ancillas only depends on the number of logical bits in this orientation and not on their arrangement. What is left to do is to pick ancilla fields such that they have a different unique ground state configuration depending on this number, and that the energy of each of these configurations is the same. Once this is accomplished, the coupling can be realized by adding additional fields to the ancillas which we will denote by $q_{i\neq0}$ . The condition that each number of logical bits in the up orientation corresponds to a single ancilla configuration can be accomplished (assuming $J^a>0$) by choosing $h_{i}^a=-J^a(2i-N)+q_i$ with $0<q_i<J_a$. The yet to be defined term $q_i$ determines an effective energy landscape depending only on the number of logical bits which are up. To match the energy landscape of \eqref{HN}, one should choose
\begin{eqnarray}
J &=& J^a, \quad h =-J^a +q_0   \nonumber \\ 
q_i &=& 
\begin{cases}
 +  J_N  +q_0  & \text{$N-i$ is odd}, \\
  -   J_N  +q_0 & \text{$N-i$ is even},
\end{cases} \label{choices}
\end{eqnarray}
with any $q_0$ which satisfies $|J_N|\ll q_0 < J^a$ and $|J_N|\ll J^a-q_0 <J^a$. The Hamiltonian \eqref{H2localemb} with coupling assignments \eqref{choices}, up to an overall constant energy offset, precisely reproduces the low energy spectrum of the Hamiltonian \eqref{HN}. For more details on this construction, see appendix 1. This is assured for the part of the spectrum with $|\mathcal{H}_N| \ll J^a$. Once $|\mathcal{H}_N| \sim J^a$ the ancillae will no longer be in their corresponding ground state and the construction breaks down. Thus the range of the spectrum which can be reproduced using the above construction depends on the maximal coupling strength of the quantum annealing device in question.  Note that the strongest field which needs to be applied for a coupler on $N$ spins is $h_N^a \approx NJ^a-q_0$.

If one only cares that the ground state is correct, and does not want to sample over a thermal distribution, than the conditions on the strength of $|J_N|$ which can be supported can be relaxed to $|J_N|< q_0 < J^a$ and $|J_N|< J^a-q_0 <J^a$. In the case where thermal sampling is desired, exactly how much less than $\mathrm{min}(J^a-q_0,q_0)$ $|J_N|$ has to be depends on both the temperature of the sampling and the accuracy desired.  The probability of a coupler having its ancillae in an excited state will in this case be roughly proportional to $\exp(-\mathrm{min}(J^a-q_0,q_0)/T)$.

The above construction can be used for example to minor embed multi-body terms in already existing architectures such as the one produced by D-wave Systems Inc, as we show in a related work on message decoding problems on the D-wave device \cite{Chancellor/Zohren2015}. In this case the fully connected graph shown in Figure \ref{fig1} must be obtained from a minor embedding \cite{Choi2011} in the Chimera graph. This is done using strong minor embedding couplings of strength $J^m$ to ``identify'' qubits. This introduces a third energy scale into the problem and one must ensure that $|J_N| \ll J^a \ll J^m$.

{\em Circuit for implementing of multi-body interactions --} In the previous section we presented an implementation which reproduces the low energy spectrum of multi-body terms using a system which only has two-body interactions. While it is possible to use minor embedding techniques to implement the above construction in already existing architectures such as D-wave, it can  be used to design a purpose-built circuit for multi-body terms. As we will show below, the fact that both logical qubits as well as ancillae have all couplings of the same strength permits a very efficient circuit implementation.

\begin{figure}
\begin{center}
\includegraphics[width=8cm]{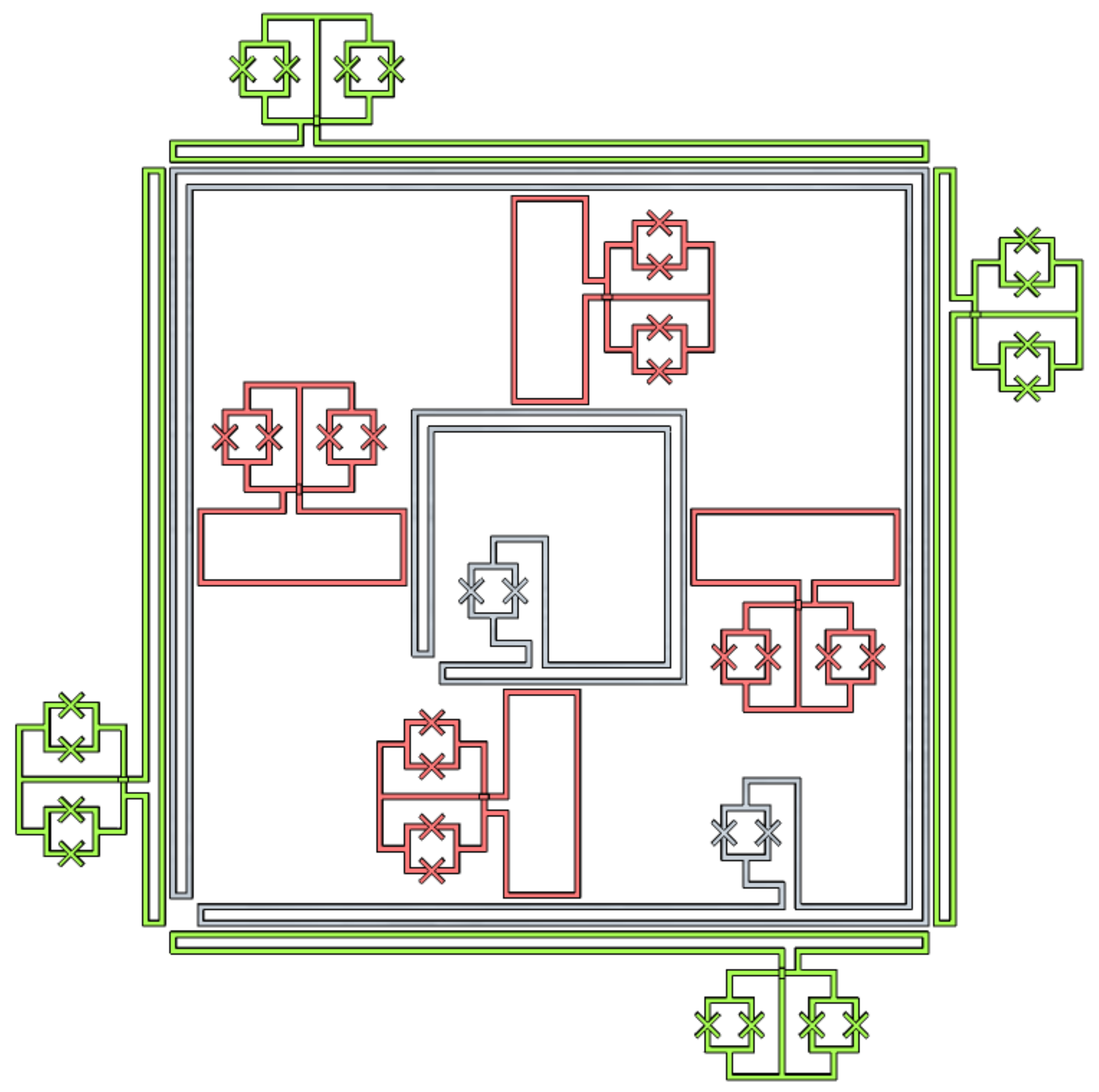}
\caption{\label{fig2} Drawing of the circuit which implements the low energy spectrum of a 4-body coupler. The 4-logical qubits appear as the 4 (green) circuits on the perimeter, with a large loop (grey) coupling them together. Further in is a set  of 4 ancilla qubits (red), which in turn themselves are coupled by an inner most loop (grey), which counteract the couplings induced from the outer loop. Note that the qubits are compound-compound Josephson junction circuits, while the couplers are simply compound. }
\end{center}
\end{figure}

In particular, in this section we present a circuit implementation of a unit cell consisting of four logical qubits coupled through a 4-body interaction. In the following section we then show how to use the unit cell to build a scalable architecture.

To implement the fully connected graph between the logical qubits as well as the coupling between the logical qubits and the ancillae, we first inductively couple all of the qubits, both logical and ancilla, to a large loop. We demonstrate later that this arrangement produces the desired graph up to the relevant order in perturbation theory. This loop is the outer loop in Figure \ref{fig2}. This coupler however also produces unwanted coupling terms between the ancillae. To cancel these off, we must add an additional loop( the inner loop in Figure \ref{fig2}) which can be biased with equal magnitude and opposite sign as the outer coupling loop to cancel  the undesirable coupling between the ancillas.

The circuit design shown in Figure \ref{fig2} requires that more than two qubits can be coupled using the same rf-SQUID coupler design as presented in \cite{van-den-Brink2005}. We demonstrate that to the same relevant order as used in the expansion in that paper, a group of qubits all inductively coupled to the same superconducting circuit realizes a fully connected 2-body graph of couplers between all of the qubits. Following \cite{van-den-Brink2005}, the energy for such a circuit with a coupler $c$ and $n$ qubits including Josephson and magnetic terms is
\begin{eqnarray}
U&=&-E_{c}\cos\phi_{c}-\sum_{i=1}^{n}E_{i}\cos\phi_{i}+ \nonumber\\
&\quad &+\frac{1}{8e^{2}}(\vec{\phi}-\vec{\phi}^{x})^{T}\mathbb{L}^{-1}(\vec{\phi}-\vec{\phi}^{x})\label{eq-U}
\end{eqnarray}
where $\vec{\phi}$ is a vector of the junction phases, $\vec{\phi^{x}}$ is the phase introduced by an external flux.  The inductance matrix $\mathbb{L}$ is given by,
\begin{equation}
\mathbb{L}=\left(\begin{array}{cccc}
L_{c} & -M_{1c} & -M_{2c} & \cdots\\
-M_{1c} & L_{1} & 0 & \cdots\\
-M_{2c} & 0 & L_{2} & \ddots\\
\vdots & \vdots & \ddots & \ddots
\end{array}\right).\label{eq-L}
\end{equation}
%

This expression can easily be inverted to obtain the term $\mathbb{L}^{-1}$ in the above expression up to $O(M^2)$. Following \cite{van-den-Brink2005}, we choose the bias fluxes such that,
\begin{equation}
\vec{\phi}^{x}=\left(\begin{array}{c}
\phi_{c}^{x}\\
\pi+(M_{1c}/L_{1})(\phi_{c}^{(0)}-\phi_{c}^{x})\\
\pi+(M_{2c}/L_{2})(\phi_{c}^{(0)}-\phi_{c}^{x})\\
\vdots
\end{array}\right), \label{eq-phix}
\end{equation}

where $\phi^{(0)}_c$ is the phase difference of the coupler due to its self induced flux, as explained in \cite{van-den-Brink2005}.   It is now worth noting that if we choose any pair of qubits $i\neq j$ and separate out only terms in \eqref{eq-U} which contain $\phi _i$ and/or $\phi _j$, these equations will be identical to those found in \cite{van-den-Brink2005}. These equations can therefore be solved independently in exactly the same way as that paper did resulting in,
\begin{equation} \label{circuitpot}
U=const+\sum_{i=1}^{n-1}\sum_{j>1}^{n}\frac{M_{ic}M_{jc}F(\phi^x_c)}{4e^{2}L_{c}L_{i}L_{j}}(\phi_{i}^{(0)}-\pi)(\phi_{j}^{(0)}-\pi),
\end{equation}
which is exactly the formula of a set of couplings realizing a fully connected graph between all of the qubits. These couplings are collectively tuneable through the function $F(\phi^x_c)$ whose exact form is not important for our purposes, but is identical to the one found in \cite{van-den-Brink2005}. They cannot however be addressed individually. For our design we desire all $L_1=L_2=...=L$ and $M_{1c}=M_{2c}=...=M$. However, as we will show, it is possible to tune for some types of imperfections in these values in for a real device.

{\em Effect of higher order mutual inductance terms}
While our calculations to demonstrate that this device will work are based on truncation of the potential energy at $O(M^2)$, in real devices one may want to make the coupling strong enough that these terms are not completely negligible. Fortunately, due to the high symmetry of our construction, spurious coupling between the logical qubits, or the ancillas, or combinations of the two  must be symmetric under permutations of the logical bits, which is the exact same symmetry as the effective coupling which the device realizes. Moreover, the  effect of non-linear coupling from higher orders in $M$ being included is effectively fixed with respect to the fields which are used to control the effective multi-body coupling, so a calibration at one coupling value will remain valid even if a different Hamiltonian is implemented. In a real device the effects of these spurious couplings need only to be measured once with appropriate compensating fields then applied. 

Because, for $J^{a}>0$ the logical states will all have the same energy, no compensation fields are necessary to deal with higher order (in $M/L$) terms arising from the outer coupling loop. The same cannot be said for the inner coupling loop however.

We now make the qualitative discussion in the previous two parargraphs quantitaitive by examining the effect of order $M^3$ terms on the four local gadget numerically. We will restrict ourselves to discussion of terms involving three distinct mutual inductances, based on the fact that terms associated with $(M_{ic})^2M_{jc}$ will come in as one and two body terms because they only involve interactions between qubit pairs which can easilty be compensated. The remaining term will be proportional to $E^{(3)}\sum_{i\neq j\neq k} \sigma^z_i \sigma^z_j \sigma^z_k$, where $|E^{(3)}| \propto M^3$ is the energy scale associated with three body terms. We now determine how strong such terms can be without causing the coupler to fail. We define failure as the case when a 'spurious' state (i.e. one in which the total magnetization of the logical and ancilla qubits is not zero) has a lower energy than the highest energy non-spurious state. Assuming that $E^{(3)}$ and the energy scale associated with the two body terms $E^{(2)}$ have the same sign, through numerical matrix analysis, we find that these higher order terms can be tolerated as long as $|\frac{E^{(3)}}{E^{(2)}}|\lesssim 8.33\%$ if $J^{a}>0$, $|J_N|=0.25\,J^a$ and $|q_0|=0.5\,J^a$, or  $|\frac{E^{(3)}}{E^{(2)}}|\lesssim 16.7\%$ if $|J_N|\ll J^a$. If the system can somehow be engineered such that $\frac{E^{(3)}}{E^{(2)}}<0$, than this number jumps to $37.5 \%$ in the case where $|J_N|=0.25\,J^a$ and $50\%$ if $|J_N|\ll J^a$. In either case, the numerical values that we have extracted demonstrate that even if the terms at the next highest perturbative order are non-negligible, or in fact relatively strong, they can be removed by applying compensating fields.


\emph{Robustness to process variability of mutual inductance}
Our realization of a multi-qubit fully connected effective 2-body graph is based on an experimentally proven design for a single coupler \cite{van-den-Brink2005}. For this reason many of the design problems have already been solved \cite{Harris2009}.
One issue which will affect our design differently is mismatches in the inductive couplings between the qubits and the loops. For a single coupler between two qubits it is not important for the mutual inductances to be matched since the coupling strength is simply proportional to a product of the mutual inductances between the two qubits and the coupler. 
For our designs on the other hand, mismatches in the mutual inductances on the outer coupling loop will lead to different coupling strengths which can be viewed as effective spurious couplings between the logical bits and/or ancillae.  

Let us first consider the effect of mismatches in mutual inductances involving the ancilla qubits. The effect of these will be twofold. Firstly, mismatched inductances will lead to imperfect cancellation of the couplings between the ancillae. If these spurious couplings are weak, they will not affect which ancillae are flipped, and just add a predictable (i.e.\ independent of which exact logical bits  are up) energy penalty in exactly the same way as the ancilla fields which are used to enforce the effective couplers. Secondly, mismatches in these couplings will mean that the couplings between each ancilla and the logical qubits will be different (although identical for a given ancilla). Again, if the mismatches are small these can be corrected for with a slight modification to the ancilla fields.

We have shown above that small mismatches in the mutual inductances between the ancilla and the coupling loops are rather benign and can be easily corrected for. Let us now consider mismatches in the inductances between the logical bits and the outer coupling loops.  These will lead firstly to effective couplings between logical bits and secondly to mismatches in the couplings between logical bits and ancillae. The former will introduce a term of the form $\sum_i \frac{\Delta M_i}{M_i} J^{a}\,\sigma^z_i\,\sum_{j \neq i} \sigma^z_j$ while the second of these terms will be of the form $\sum_i \frac{\Delta M_i}{M_i} J^{a}\,\sigma^z_i\,\sum_{j} \sigma^z_{j,a}$. If we assume that the mismatches in the mutual inductances are small enough that they do not change the order in which the ancillae flip, then the ancillae states just represent a count of the logical spins which are up and the second of these terms can be written as $-\sum_i \frac{\Delta M_i}{M_i} J^{a}\,\mathrm{sgn}(J_{a})\,\sigma^z_i\,(\sum_{j \neq i} \sigma^z_{j,a}+\sigma^z_i)$. Therefore, as long as $J^{a}>0$, the effect of such mismatches will only be to introduce an irrelevant constant energy offset. We can therefore conclude that if the coupling $J_N$ is anti-ferromagnetic, then via adjustment of the local fields on the ancillae, \emph{all mutual inductance mismatches below a certain threshold can be corrected for with existing controls}.

This correction threshold is the point at which the order which the ancillae flip deviates from the case where the inductances are all matched. While finding the exact value of mutual inductance mismatches at which this happens depends on the specific errors in a detailed way, we can make simple arguments to find lower bounds for this threshold. We first note that the maximum energy shift of any state caused by a mispecification on the mutual inductance between an ancilla and the inner loop cannot be more than $\frac{\Delta M^{inner}_{a,i}}{M^{inner}_{a,i}}\,J^{a}\,(N-1)$. Similarly for coupling with the outer loop, the energy shift cannot be more than $\frac{\Delta M_{(a),i}}{M^{inner}_{(a),i}}\,J^{a}\,(2\,N-1)$. 
In addition to these shifts, the spectrum will be modified by correction terms which must be added to the Hamiltonian. Each of these corrections must also be smaller than the maximum energy shift. Furthermore, due to symmetry, all logical states are affected in the same way by any mismatch involving the outer coupling loop, therefore only mismatches on the inner loop will require corrective modification of the fields. By summing up all possible shifts and corrections, and comparing to the minimum energy difference between a state where the ancillae correctly map the configuration of the logical spins and states where they do not, we conclude that mutual inductance mismatches will certainly be correctable as long as 
\begin{align}
\mathrm{min}(||J^a|-q_0|-|J_N|,|q_0|-|J_N|)>\nonumber \\
\mathrm{max}(\{\left|\frac{\Delta M}{M}\right|\})\,|J^{a}|\,(2\,N\,(N-1)+2\,N\,(2\,N-1)).
\end{align}
Where $\{\left|\frac{\Delta M}{M}\right|\}$ indicates the set of all mutual inductance mismatches. While simple, this bound is not necessarily tight- there may be many cases where this bound is not met, and yet it is still feasible to correct for inductance mismatches. This Nevertheless demonstrates that for the four local device, errors of at least $\left|\frac{\Delta M}{M}\right|\lesssim 0.63 \%$ can be tolerated if we choose $|q_0|=\frac{1}{2}|J_a|$ and $|J_N|\ll J^a$.

\begin{figure}
\begin{centering}
\includegraphics[width=6cm]{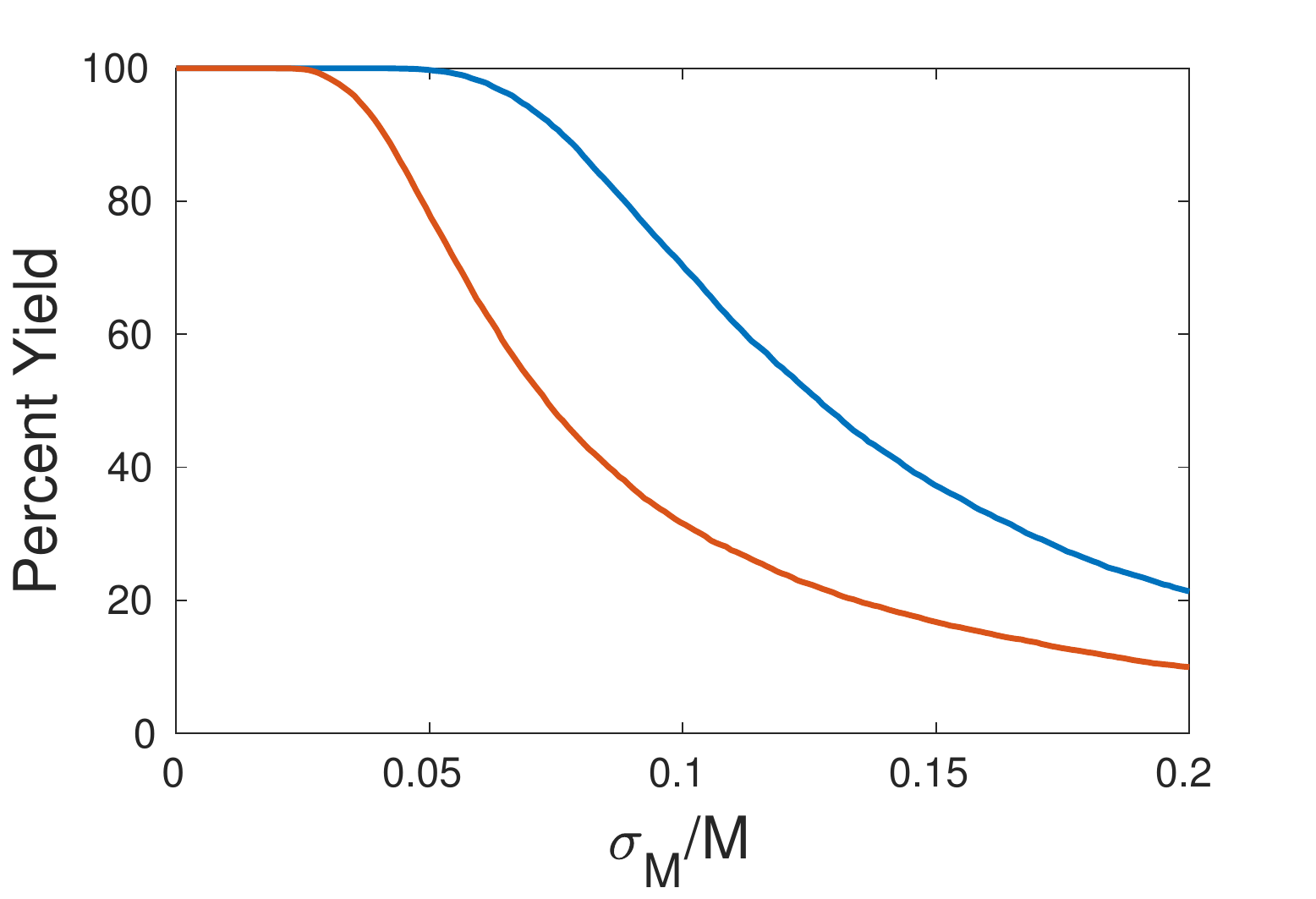}
\par\end{centering}
\caption{\label{fig:yield_rate} Yield rates for four local coupler circuits with $|q_0|=\frac{1}{2}|J_a|$ versus deviation of $M$, $\frac{\sigma_M}{M}$. The blue curve is for $|J_N|\ll J^a$ while the brown is for $|J_N|=0.25\,J^a$.}
\end{figure}

As Fig. \ref{fig:yield_rate} illustrates, this bound is overly pessimistic. This figure shows the modelled yield rate as a function of the normalized standard deviation in mutual inductance between the coupling loops and the qubits and ancillae for a four-body coupler circuit with $|q_0|=0.5\,|J_a|$. Here yield is defined as the fraction of modelled circuits for which correction by tuning of the ancilla local fields allows the implementation of Hamiltonian (1) to be successfully implemented. Failure in this case is defined as a spurious state (i.e. one in which the total magnetisation is non-zero, see Table 1) occurring at lower energy than any non-spurious state. We assume that mutual inductance errors follow independent Gaussian distributions. We find that for $|J_N|\ll J^a$, the circuit will almost always be correctable even if the standard deviation in $M$ is as large as around $5 \%$. For $|J_N|=0.25\,J^a$, it will almost always work if the standard deviation in $M$ is as large as around $2.5 \%$.  Of all $10,000$ instances of mismatch rates tested, the lowest values of $\frac{\sigma_M}{M}$ where failure was observed (i.e. a spurious state had a lower energy than a non-spurious state) were $0.0420$ for $|J_N|\ll J^a$ and $0.0193$ for $|J_N|=0.25\,J^a$.

In practice the largest source of uncertainty in the mutual inductance occurs at the design stage since finite element methods typically are accurate to around $10\%$. Fabrication of test structures enable this uncertainty to be corrected to around $1\%$ \cite{Topygo2015}. Furthermore if necessary the mutual inductance can be adaptively corrected by applying a flux to a SQUID loop between the qubits or ancillae on one hand and the coupling loops on the other, following the approach of \cite{Harris2009}.


The requirements on the ancilla are much less strict than those on the logical qubits. They do not need to be read out at the end of the annealing run.  Furthermore, there is no reason to which the ancillae need to be run on the same annealing schedule as the logical qubits. This presents a significant advantage in that the annealing schedule in this architecture naturally breaks down into a path in a space defined by two parameters. Adjusting the annealing parameter (i.e. the ratio between the longitudinal and transverse fields) on the ancillas adjusts how strongly the multi-body constraints are enforced. The annealing parameter on the logical bits on the other hand acts analogously to the way it does in the standard 2-body transverse field Ising model.

This natural breakdown into two independent annealing parameters allows for several interesting possibilities. For one this provides a natural testbed for an optimizable annealing schedule. Secondly, by introducing a control element which acts non-deterministically, a device could be designed which anneals  on a different schedule each run.  If a pathological region (for example one with a very small gap) exists in the annealing trajectory space, such a protocol may be able to avoid this region during some of the runs, thereby increasing the robustness of the annealing algorithm. A quantitative analysis of such two-parameter annealing schedules goes beyond the scope of this paper and would be a rich area for further analysis.

The techniques we propose here could be used to construct an effective three local coupler with three ancillae. This can be done more efficiently however using a gadget proposed independently in \cite{Chancellor/Zohren2015,Lechnernew} which also has symmetry under permutation of the logical bits. Because the circuit implementation of this simpler design for three bit couplers only differs slightly from the general design, we reserve discussion of this to section 2 of the supplemental material.
\begin{figure}[t]
\begin{center}
\includegraphics[width=8cm]{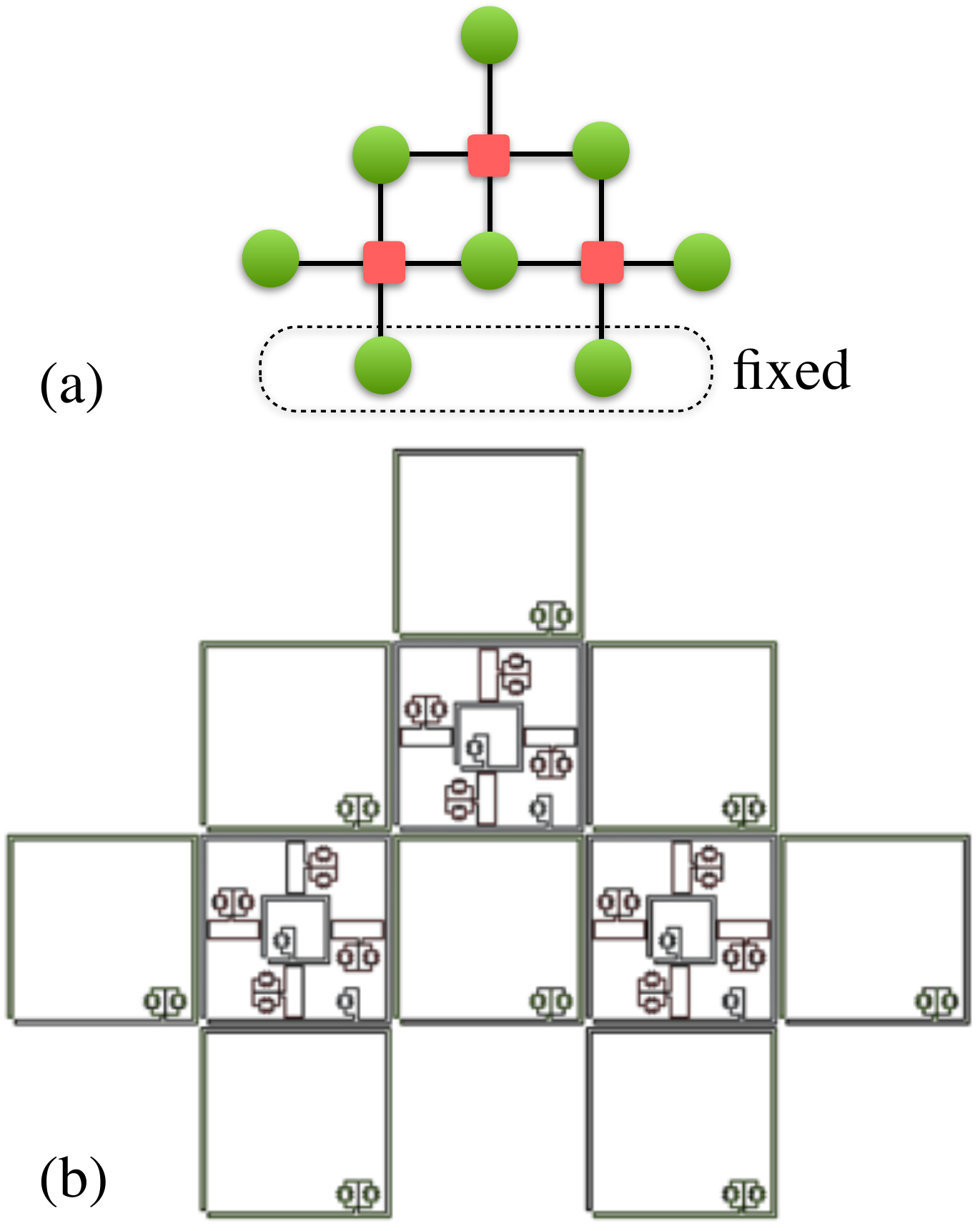}
\caption{\label{fig3} (a) Schematically illustration of the embedding technique proposed in \cite{Lechner2015}. Shown is an embedding for $M=3$ logical qubits whose interactions are mapped onto $K=6$ physical qubits (green circles). Besides the six physical qubits encoding the interactions, there are an additional two physical qubits which enforce the boundary condition. The physical qubits are coupled using four-body interactions as illustrated by the plaquette (red squares). (b) A physical circuit implementation of the above embedding using our circuit design of a four-body term as presented in the previous section. The four qubits surrounding the four-body circuit are extended to couple to neighboring four-body circuits. This presents a concrete circuit design for a scalable quantum annealing architecture using superconducting qubits.}
\end{center}
\end{figure}

{\em Applications to construct a scalable architecture -- } We now show how our above circuit design can be used to construct a scalable architecture using a recently proposed embedding techniques \cite{Lechner2015} which maps $M$ logical qubits with full connectivity to $K=M(M-1)/2$ physical qubits which have 4-body nearest neighbour interactions (see Figure \ref{fig3} (a)). More concretely, in \cite{Lechner2015} a model with 2-body interactions but arbitrary connectivity is considered
\beq
\mathcal{H}= \sum_{i =1}^M  \sum_{j<i }   J_{ij } \sigma^{z}_i \sigma^{z}_j,  
\eeq
where qubits are on a fully connected graph and we have left out the magnetic fields for simplicity. We see that we have $M(M-1)/2$ degrees of freedom in the assignment of the couplings $J_{ij}$.  In the embedding proposed in \cite{Lechner2015} the system is mapped to a system with $K=M(M-1)/2$ physical qubits $\tilde{\sigma}^z_i$ arranged in a pattern as shown in Figure \ref{fig3} (a), where the $K$ original couplings $J_{ij}$ map on the $K$ magnetic fields $\tilde{b}_i$ of the physical qubits. The original qubits are associated with $M$ four-body interactions, shown as plaquette in Figure \ref{fig3} (a). The resulting Hamiltonian is 
\beq
\mathcal{H}= \sum_{i =1}^K   \tilde{b}_i \tilde{\sigma}^{z}_i  - C \!\!\!\! \sum_{(i,j,k,l)\in\mathrm{plaquettes}}  \tilde{\sigma}^{z}_i \tilde{\sigma}^{z}_j\tilde{\sigma}^{z}_k\tilde{\sigma}^{z}_l
\eeq
In Figure \ref{fig3} (b) we illustrate how such an architecture can be physically implemented using our circuit design of a four-body term. The resulting architecture is scalable in the sense that adding a new logical qubit simply amounts to adding a new row at the bottom. The final row is also used for readout of the values of the logical qubits as explained in \cite{Lechner2015}. 

Combining the embedding proposed in \cite{Lechner2015} with the circuit design for four-body interactions between rf-SQUIDs as described in the previous sections provides a concrete implementation for a scalable architecture allowing for two-body interactions of arbitrary connectivity. Interactions between rf-SQUIDs are mediated by the simple circuit design described above which involves at maximum  couplings between five loops. 

One major concern with rf-SQUID systems is the presence of noise which couples inductively into the system. The primary goal of the architecture proposed in \cite{Lechner2015}, is not to reduce such noise, but rather to provide an alternative method of embedding problems to the traditional minor embedding approach \cite{Choi2011, Venturelli2014}. Fortunately however, most improvements in fabrication and other techniques which would benefit devices with architectures which require minor embedding to map highly connected graphs would also benefit systems where problems are embedded using the methods of  \cite{Lechner2015}. 

Whether a circuit implementation using the architecture proposed here is more or less noisy than ones using other rf-SQUID architectures is likely to depend on details of the physical implementation. In particular, in designing the value of the circulating current in the rf SQUID qubit annealer there is a trade-off between coupling and coherence. Smaller values of circulating current lead to longer coherence lifetimes (due to the lower flux noise) and will require higher values of qubit-coupler mutual inductance which will in turn be less susceptible to process variability. One advantage of our method however is that all of the qubit circuits are planar, which may lead to a relatively simpler fabrication process, and therefore more flexibility in terms of making modifications to reduce noise.

\section{Methods}

Calculations were first performed by hand and then verified by a simple Matlab scripts containing less than 150 lines of code in total, including subroutines.  For the yield calculations, we generated error Hamiltonians $H_{error}$ with independent Gaussianly distributed couplings corresponding to mis-specifications of each mutual inductance. We then examined the eigenstates of $H_{tot}=H_{Coupler}+\frac{\sigma_M}{M}H_{error}$ and found the value of $\frac{\sigma_M}{M}$ where spurious state energies crossed logical state energies. Note that this techniques ignores higher order interaction terms from these mismatches, which would have an effect of order $(\frac{\sigma_M}{M})^2$. A similar technique is used to calculate the strength of higher order terms which can be compensated. While our code could be quickly and easily reproduced by anyone with basic proficiency in programming, it is also available from the authors upon request.

\section{Discussion}

 Overcoming the physical limitations on connectivity and multi-body interactions of the underlying Ising spin system in hardware implementations of quantum annealing devices is a major challenge. In this work we present a method of effectively implementing a Hamiltonian with multi-body terms by reproducing its low energy spectrum using a Hamiltonian which only involves two-body interactions and a number of ancilla qubits. While this construction can be used as a minor-embedding technique for existing quantum annealing architectures (as we explore in a related work \cite{Chancellor/Zohren2015}), the major result of this work is an efficient circuit design of the construction using superconducting flux qubits as well as calculations which demonstrate that this circuit is robust to realistic design imperfections. Having the possibility of inductively coupling a number of qubits with all-to-all couplings of equal strength using a single coupling loop enables us to implement the multi-body terms using a very efficient circuit design. In the last section we show concretely how this circuit can be used as a unit cell for a scalable architecture using a recent embedding technique \cite{Lechner2015} which encodes logical qubits of arbitrary connectivity using physical qubits with nearest-neighbour four-body interactions. 

As a final remark, let us mention that here we have focused on circuit based implementations of multi-body terms in the $z$-basis, i.e.\ $\sigma_1^z\ldots\sigma_N^z$. This is natural, since the optimization problem is encoded in the $z$-basis. However, one can in principle also consider similar constructions in the basis of the transverse magnetic field, i.e\ $\sigma_1^x\ldots\sigma_N^x$, thereby enabling implementation of non-stoquastic Hamiltonians \cite{Brayvi} involving multi-body terms. We leave such an analysis for future work.

{\em Note added --}  After a preprint of this letter appeared on arXiv, a related paper \cite{Lechnernew} was released which proposes an implementation of multi-body terms using Transmon qubits. Further, a revision of \cite{Andrea} was released which added an abstract construction of multi-body terms using a chaining of 3-body terms. While no physical implementation is discussed in this paper, the abstract formalism relates to that of \cite{Lechnernew} and could be implemented in a similar manner.



\section{Acknowledgements}

\begin{acknowledgments}
The authors would like to thank Simon Benjamin and Stephen Roberts for discussions. NC and PAW were supported by EPSRC (grant refs: EP/K004506/1 and EP/H005544/1) and Lockheed Martin . S.Z. acknowledges support by Nokia Technologies, Lockheed Martin and the University of Oxford through the Quantum Optimisation and Machine Learning (QuOpaL) Project.

\end{acknowledgments}


\section{Contributions}

The authors all contributed to the development of the idea, and all wrote the paper together. SZ created the diagrammatic figures, while NC produced all numerical data and created the yield rate figure.

\section{Competing Interests}

The authors have jointly filed for a (UK) patent for the coupler design described in this document.

\bibliography{}{}

\section{Appendix 1: Construction of the Hamiltonian}

In this appendix we explain in a detailed way how our construction works, including discussion about how we arrived at it. We will first discuss the role which symmetry plays in our design. We will then explain how this symmetry can be exploited to construct a Hamiltonian in which the ancillae `count' the number of logical spins in the up configuration. For this construction, we first discuss how to construct a combination of couplers and fields for which the ground state orientation of each ancilla flips when a given number of logical bits are in the up orientation. The counting is achieved by adding a number of such ancillae equal to the number of logical spins. Following this construction we explain how additional terms can be added to construct a `neutral' Hamiltonian, in which the lowest energy state given a fixed configuration of the logical bits will always have the same energy. Finally, we explain how additional biases can be added to the ancillas to transform this neutral Hamiltonian into the coupler defined in Eq. (1) of the main document.

\paragraph{Symmetry}

The first thing to note about the Hamiltonian in Eq. (3) of the main document is that it is symmetric under permutation of any of the logical qubits. This symmetry guarantees that the total energy depends only on the number of logical bits in the `up' orientation and not on their arrangement. This permutation symmetry is also shared by the coupler Hamiltonian, Eq. (1) of the main document. In fact, as discussed in \cite{Chancellor/Zohren2015} our construction here can be easily generalized to any Hamiltonian with this symmetry. There is another more subtle result of this symmetry as well: the ground state configuration of the ancillae must also only depend on the number which are `up' rather than their configuration.

\paragraph{Counting}

What we want to do is to produce a Hamiltonian for which the ground state ancilla configurations `count' the number logical bits which are in the `up' orientation. More precisely what we mean by this is that there is a unique state of the ancillae which is both the same for all arrangements of the logical bits and which is distinct for each number of logical bits which are `up'. The first of these requirements is automatically guaranteed by symmetry. The second of these needs to be achieved more carefully.

\begin{figure}[t]
\begin{center}
\includegraphics[width=6cm]{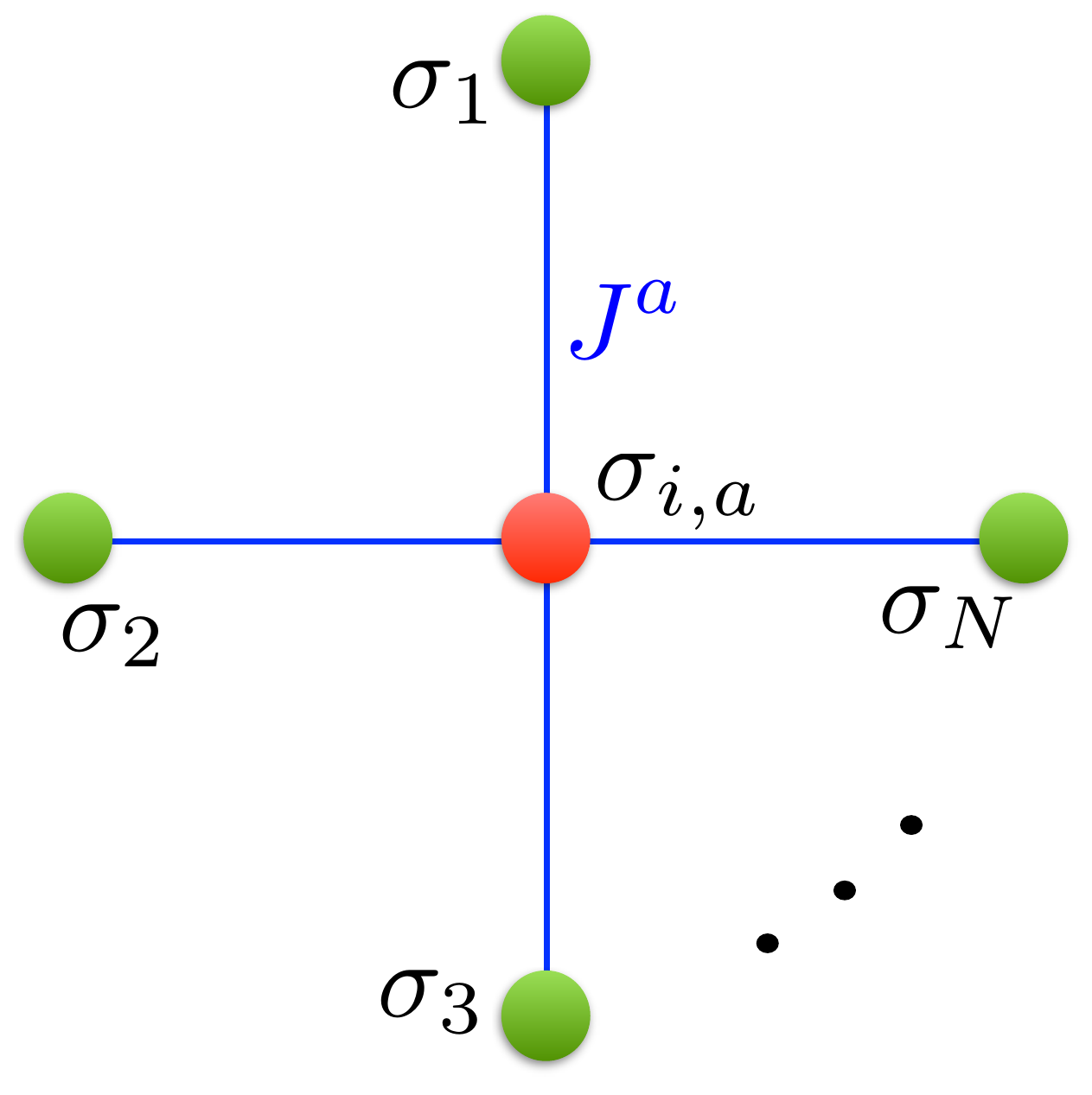}
\caption{\label{fig1anc}Single ancilla coupled to all logical qubits equally.}
\end{center}
\end{figure}

Since the ancillae are not directly mutually coupled, we can consider each ancilla separately as shown in Fig. \ref{fig1anc}. In addition, until later in the analysis, we set the the local fields on each of the logical qubits to zero. There are two ways in which the ancilla state can affect the total energy: firstly through its coupling to the logical bits (which must all be identical by symmetry) and secondly  through its own local field $h^a_i$. For simplicity, and because it will be desirable in our circuit design, we set all the ancilla-logical qubit couplers to be equal to $ J_a$. 

The energy of this system is $E_{a,i}=\sigma^z_{i,a} (J_a N_{up}+h_{i}^{a})$, where $N_{up}$ is the number of logical bits in the `up' orientation. From this observation, we notice that the ground state orientation of the ancilla will flip when $J_a N_{up}+h_{i}^{a}$ changes sign. Therefore if we desire to have an ancilla flip when $N'_{up}$ logical bits are up, then we must choose $-J_a (N'_{up}-1)>h_{i}^{a}>-J_a N'_{up}$. Therefore, a set of $N$ ancillas, each fulfilling this condition for a different  $0\le N'_{up} \le N$ will `count' the number of logical up spins in the sense that the number of ancillae which are down will match the number of logical bits which are up, and that the ancillae will flip in a set order. More specifically we set each of these ancilla fields to $h_{i}^a=-J^a(2i-N)+q_0$, where $0<q_0<J^{a}$  is a free parameter. 

\paragraph{Neutral Hamiltonian}

We will later add a bias to this to implement the coupler Hamiltonian, but first we want to construct a `neutral' Hamiltonian, where the total energy with the ancillae in the ground state is the same for all logical bit configurations. We now consider the complete coupler Hamiltonian from Fig. 1 of the main document and described by Eq. (3) of the main document. The four separate terms of Eq. (3) of the main document can be re-expressed in terms of $N_{up}$ using the following relationships

\begin{eqnarray}
\sum_{i=1}^N\sigma^z_i \!&=&2 \! N_{up}-N, \\
\sum_{i=2}^N \sum_{j =1}^{i-1} \sigma^{z}_i \sigma^{z}_j  \! &=& \! \frac{1}{2}(2\,N_{up}-N)^2-\frac{1}{2}N, \\
\sum_{i=1}^N \sum_{j=1}^N \sigma^{z}_i \sigma^{z}_{j,a} \! &=& \! -(2\,N_{up}-N)^2-2\,N_{up}, \\
 \sum_i  h_i \sigma^{z}_i \! &=& \!\!
J_a(-2N_{up}^2+2 N N_{up}+2 N_{up}-N)\! - \nonumber \\
&& \quad -q_0\,(2\,N_{up}-N),
\end{eqnarray}
Hence Eq. (3) of the main document becomes 
\begin{eqnarray}
\mathcal{H}_N^{(2)} &=&  h \nu +J (\frac{1}{2}\nu^2 -\frac{1}{2}\,N)+ \nonumber \\ 
&& \quad +J_a\,(-\frac{1}{2}\nu^2+\nu-\frac{1}{2}N^2-N)-q_0\nu, 
\end{eqnarray}
where we have defined $\nu=2\,N_{up}-N$ for compactness. From this formula, it is clear that, when the ancillae are in the ground state, the energy wil be intependent of the number of logical up spins if we choose $J=J_a$ and $h=-J^a+q_0$.

\paragraph{Coupler Biasing}

The final step in producing a coupler is to bias the ancillae such that the states where $N_{up}$ is odd have different energies than those for which $N_{up}$ is even. This step is simply accomplished by adding an alternating bias as illustrated in Eq. (4) of the main document. This bias must be weak enough to not change the ground state configuration of the ancillae, meaning that $|J_N| < q_0 < J^a$ and $|J_N| < (J^a-q_0) < J^a$.

\section{Appendix 2: Specialized 3-local design}

While the methods we propose can produce three-local coupling using three ancillae the gadget proposed independently in \cite{Chancellor/Zohren2015,Lechnernew} is more efficient in that it requires only a single ancilla. This gadget has symmetry with respect to permutations of the logical bits as the general N-local one we propose here, and also has all logical bits coupled to the ancilla. As we demonstrate, this means that we can take advantage of the same circuit design tricks. Before discussing the circuit design, let us first briefly introduce the more efficient three local gadget.

As discussed previously, this Hamiltonian has the same structure as our N-local gadget, but with only a single ancilla, its Hamiltonian takes the form,

\begin{eqnarray}
\mathcal{H}_3^{(2)}&=& J \sum_{i=1}^3 \sum_{j =1}^{i-1} \sigma^{z}_i \sigma^{z}_j + h \sum_{i=1}^3  \sigma^{z}_i + \nonumber \\ 
&& \quad + J^a \sum_{i=1}^3  \sigma^{z}_i \sigma^{z}_{a} +    h_a \sigma^{z}_{a}. 
\label{H2local3}
\end{eqnarray}

To reproduce the spectrum of a three local coupler, one then chooses $h=J_N$, $J_a=2J>h$, and $h_a=2h$. To perform sampling tasks, one must choose $2J \gg g$.

\begin{figure}[t]
\begin{center}
\includegraphics[width=8cm]{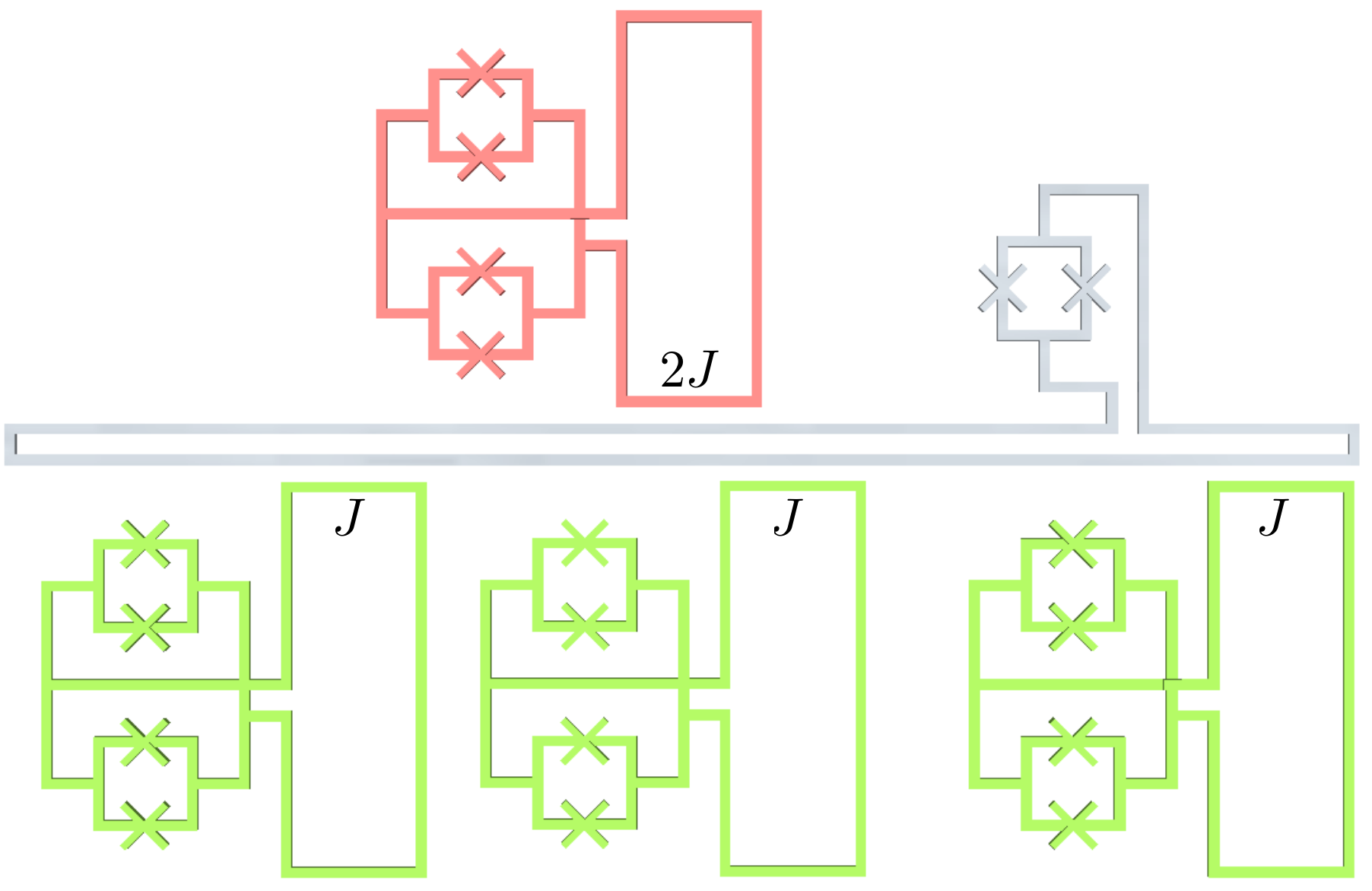}
\caption{\label{fig3circuit} Shown is the circuit to implement of the more efficient Hamiltonian for $N=3$, as given in \eqref{H2local3}. The coupling loop couples to the logical qubits and the ancilla with different strengths, namely $J$ and $2J$ respectively.}
\end{center}
\end{figure}

Because this Hamiltonian only has a single ancilla, there is no requirement to remove coupling between the ancillae. However, there is an additional complication which we did not have previously, $J_a \neq J$.  However, such coupling can easily be implemented with a single loop which couples all of the logical qubits, and the ancilla, but with the mutual inductance between the ancilla and this loop twice as strong as all of the others, by Eq. (5) of the main document the coupling to the ancilla will only be half as strong. This circuit is depicted in Fig. \ref{fig3circuit}

This three local circuit design has many potential applications, for one it can be used to construct an alternative version of the coupling scheme proposed in \cite{Lechnernew}, but based on flux qubits rather than transmons.

\bibliography{}{}

\end{document}